# RF-Sputtering Deposition of $Nd_{1-x}Sr_xCoO_3$ Oriented Thin Films

*Lorenzo Malavasi[1,*], Carla Sanna[2], Nathascia Lampis[2], Alessandra Geddo Lehmann[2], Cristina Tealdi[1], Maria Cristina Mozzati[3], and Giorgio Flor[1]*

[1]Dipartimento di Chimica Fisica "M. Rolla", INSTM, Università di Pavia, V.le Taramelli 16, I-27100, Pavia, Italy.

[2]Dipartimento di Fisica – Università di Cagliari - Cittadella Universitaria St. Pr.le Monserrato-Sestu km. 0.700, I-09042 Monserrato (Ca), Italy.

[3]CNISM, Unità di Pavia and Dipartimento di Fisica "A. Volta", Università di Pavia, Via Bassi 6, I-27100, Pavia, Italy.

In this paper we reported, to the best of our knowledge, the first deposition of highly oriented thin films (with thickness of about 90 nm) of $NdCoO_3$ and $Nd_{0.8}Sr_{0.2}CoO_3$ cobaltites on single-crystalline STO and LAO substrates.     Our investigation has shown that highly oriented single phase thin films of NCO and NSCO can be successfully deposited by means of rf-sputtering if the substrates is heated at high temperatures (700°C); lower substrate temperature has shown to lead to multi-phase materials with a low crystallinity degree . LAO substrate showed to give origin to a prefect match of the out-of-plane lattice constant of the NSCO target material.

*Corresponding Author: Dr. Lorenzo Malavasi, Dipartimento di Chimica Fisica "M. Rolla", INSTM, Università di Pavia, V.le Taramelli 16, I-27100, Pavia, Italy. Tel: +39-(0)382-987921 - Fax: +39-(0)382-987575 - E-mail: lorenzo.malavasi@unipv.it

# Introduction

Cobalt containing perovskites are highly attractive materials which are object of intense and actual research from several different areas such as sensors research, magnetism and fuel-cell technology [1-3]. For this reason, the lattice, electronic and magnetic properties of compounds such as $LaCoO_3$ have been subject of continuing interest since the 1950s [4-6]. More recently, an interesting debate has been opened regarding the actual spin-state of Co in lanthanum cobaltites [7]. In fact, it has been shown that in the range from 5 to 1000 K the Co ions pass through three different spin-states (low-spin LS, intermediate-spin IS, and high-spin HS) which are intimately connected to the structural as well as internal parameters such as the metal-oxygen bond lengths.

It is also well known that rare-earth cobaltites ($RECoO_3$) present certain similarities with the manganites in terms of transport and magnetic properties such as the doping induced ferromagnetism and metallic conductivity [8]. However, transition temperatures from insulating to metallic (IM) state and from paramagnetic to ferromagnetic (PF) order are not so closely coupled as in the manganite perovskites [9]. Among $RECoO_3$ compounds, $LaCoO_3$ and $NdCoO_3$ have been object of a deep investigation both as pure compounds and considering the role of divalent dopant (Sr) concentration [8,10-12]. Recent results of $^{59}Co$ NMR on the $Nd_{1-x}Sr_xCoO_3$ solid solution [10], for example, have pointed out a phase separation *scenario* as the main ground state for doped cobaltites which, in fact, behave as "bad metal" at high temperatures and high doping-levels, where, on the opposite, the analogous manganites are good metals. The additional "degree of freedom" present in the cobaltites, with respect to manganites, is the possibility to find different spin-states which depends mainly on the chemical pressure effect induced by the cation dimension on the A-site of the perovskite structure. The spin-state, in turn, strongly influences the transport and magnetic properties of the materials. A general gradient approximation (GGA) study due to Knižek [13] demonstrated that the relative stability of IS and LS depends on the Co-O distances and angles with longer bonds and more open Co-O-Co angle favouring the IS state.

It is clear that the internal parameters such as the bond lengths and angles play a major role in the defining the properties of the cobaltites, in a similar or even more important way as in the mangnanites. Among the possible ways of tuning these parameters we have already put in prominence the cation replacement, i.e. the application of a chemical pressure; in addition we may recall the application of an hydrostatic pressure or the variation of the oxygen content. However, another very efficient way of changing the "pressure" on the system is to induce a degree of strain by depositing epitaxial thin films on specifically oriented substrates. This has shown to be a very intriguing and elegant manner of finely changing the manganites properties [14-16].

Our aim is then to try to apply this last aspect on the cobaltite compounds. Among the current literature, to the best of our knowledge, there are no papers dealing with the synthesis of epitaxial, oriented, thin films of any cobaltite. We have already reported the preparation of un-oriented thin films of pure $NdCoO_3$ for sensing applications towards CO [17]. Here we try to extend our method to the deposition of epitaxial $NdCoO_3$ (NCO) and $Nd_{0.8}Sr_{0.2}CoO_3$ (NSCO) films on different substrates, namely $SrTiO_3$ (001) and $LaAlO_3$ (100).

This paper reports the preliminary results of this on-going work comprising the X-ray characterization of the prepared thin films.

## Experimental Section

Powder samples of $NdCoO_3$, and $Nd_{0.8}Sr_{0.2}CoO_3$ have been prepared by conventional solid state reaction from the proper stoichiometric amount of $Nd_2O_3$, $Co_3O_4$, and $SrCO_3$ (all Aldrich ≥99,9%) by repeated grinding and firing for 24 h at 900-1050 °C.

Thin films were deposited onto single-crystalline $SrTiO_3$ (001), and $LaAlO_3$ (100) (Mateck) by means of off-axis rf-magnetron sputtering (Rial Vacuum). The gas composition in the sputtering chamber was argon and oxygen (16:1) with a total pressure of $4 \times 10^{-6}$ bar. The substrate was heated at 700°C and rotated during deposition. The rf power was set to 150 W. After the deposition the films were annealed at 900°C in pure oxygen for 30 minutes. The chemical composition of starting powders and thin films was checked by means of electron microprobe analysis (EMPA) which confirmed their correct cation ratio.

X-ray diffraction (XRD) patterns and X-ray reflectivity (XRR) measurements were acquired on a "Bruker D8 Advance" and "Bruker D8 Discover" diffractometers equipped with a Cu anode.

## Results and Discussion

Figure 1 reports the refined X-ray diffraction pattern of the NCO target material. The pattern can be perfectly refined considering an orthorhombic unit cell (space group n. 62, *Pnma*) with lattice constants $a = 5.3380(2)$, $b = 7.5545(4)$, and $c = 5.3509(2)$. For the NSCO target the final structural parameters obtained from the refinement are: $a = 5.3470(1)$, $b = 7.5857(4)$, and $c = 5.3757(2)$. Doping the Nd-site with 20% of Sr induces a slight (~1%) cell expansion from 53.945(5) Å$^3$ to 54.510(5) Å$^3$. This relatively small expansion, with respect to the difference in the ionic radii (for the same coordination, 12) between Nd$^{3+}$ (1.27 Å) and Sr$^{2+}$ (1.44 Å) is due to the concomitant oxidation of Co ions. In fact, we expect that the Sr-doping will increase the hole concentration according to the following equilibria which take into account the cation replacement and the compensation of oxygen vacancies with external oxygen:

$$2NdCoO_3 + 2SrO \Leftrightarrow 2Sr'_{Nd} + V_O^{\bullet\bullet} + Nd_2O_3 + 5O_O^x \qquad (1)$$

$$2V_O^{\bullet\bullet} + O_2 \Leftrightarrow 4h^{\bullet} + 2O_O^x \qquad (2)$$

Starting from the NCO and NSCO targets we deposited a series of thin films on single-crystalline substrates. We chose to grow the films on cubic STO (001) and LAO (001). The choice has been done considering the lattice constants of the target materials with respect to the parameters of the substrate. In particular, the cubic axis for STO is 3.90 Å while for LAO it is 3.82 Å. Considering a pseudo-cubic cell for the NCO and NSCO samples, we may expect an *average* parameter around 3.78 Å for the NCO and 3.79 Å for the NSCO. So, in principle we should be able to look at the role of substrate nature on the growth of the films and on the physical properties induced by possible modulation of the lattice parameters (*i.e.* strain).

The deposited films have been first characterized through X-ray reflectivity. Figure 2 reports, as an example, the XRR spectrum for the pure NCO thin film deposited on STO for about 30 minutes. In the inset it is highlighted a small part of the spectrum in order to put in evidence the Kiessig fringes which originate from the interference between successive layers of the film. The separation between two successive maxima in the curve is a direct and reliable measurement of the thin film thickness, *t*. However, we calculated the *t* value from a fit of the experimental curve with a model by means of the LEPTOS software (Bruxer AXS). It turned out that the film thickness is around 90 nm, thus indicating a deposition rate of ~3 nm/min. All the films considered in this work showed to be, within the experimental error, of the same thickness.

Figure 3 reports the X-ray diffraction patterns for the NSCO thin films measured immediately after the deposition ("as-dep", solid line), and after the oxygen annealing ("annld", dashed-line), respectively; in addition, vertical grey bars represents the STO peaks (also labelled in the Figure). First of all we note that the thin films are highly oriented since they clearly present the only peaks related to those of the substrate. Before the annealing treatment the FWHM of the peak at around 48° is 0.352° and slightly reduces to 0.327° after the thermal treatment thus suggesting the crystallinity of the films is not significantly improved by this relatively short treatment.

Concerning the thin films orientation we can not be conclusive based on these XRD data. In fact, considering the pseudo-cubic cell parameters deriving from the orthorhombic ones and assuming that the thin films adopt the *Pnma* crystal structure of the target material, two orientations of the films, with respect to the substrate, can be found: i) [010]-orientation, and thus the peaks in Figure 3 correspond to the (020) and (040) reflections of the cobaltite, and ii) [101]-orientation, with peaks in the pattern corresponding to the (101) and (202) cobaltite planes. This point is illustrated in more details in Figure 4, where, for a wider specular scan, the two sets of Miller indexes for the two different epitaxial orientations of the NCSO films are indicated. Finally, we may not exclude that the substrate induced the growth of more symmetric films with respect to the target materials. In particular, cubic or, more probable, tetragonal symmetry may not be ruled out based on these data. Further XRD experiments will be used in order to determine the in-plane lattice parameters and, consequently, exactly define the thin films orientation.

From the X-ray patterns of Figure 4 we calculated the out-of-plane parameter for the NSCO thin film before and after the annealing treatment. In the first case the out-of-plane coordinate is 3.781(1) Å while after that treatment the parameter slightly contracts to 3.771(1) Å. This is most probably due to the partial oxidation of the cobalt ions with the creation of smaller oxidised species such as $Co^{4+}$.

Figure 5 reports the XRD pattern for the 90 nm NCO thin film deposited on STO (001). As can be appreciated, pure NCO grows with a high degree of orientation in analogous way as NSCO did. The out-of-plane parameter for NCO, as calculated from the position of the peaks, is 3.766(1) Å. This has to be compared with 3.771(1) Å of NSCO. The difference between these parameters is a bit smaller with respect to the difference in the bulk cell which may originate from the effect induced by the substrate. In the inset it is presented the comparison between the two thin films in a limited region of the XRD pattern.

Finally, the influence of substrate temperature during thin films deposition is put in prominence through Figure 6. Here, it is reported the XRD pattern of a NCO thin film deposited heating the substrate to 400°C instead of 700°C, as done for the other films considered in this work, while keeping all the other deposition parameters constant. As can be appreciated, beside the cobaltite phase analogue to the previous samples other intense peaks appear in the pattern (marked with an asterisk). At present we are not able to undoubtedly associate these peaks to a precise phase. The peak located at about 38° might be the (102) and (201) reflections of the orthorhombic cobaltite, thus indicating that in order to obtain full oriented films the deposition has to occur heating the substrate at high temperatures. Anyway, peak at around 65.5° can not be related to the cobaltite structure. So, another possibility is that these peaks originate from a second phase (which is not, however, any of the simple metal oxides of Co or Nd) of still uncertain composition. We also remark that the FWHM of the NCO peaks are around 0.55° which strongly indicates that higher deposition temperatures are effective in improving the thin film crystallinity.

To conclude, Figure 7 reports the XRD pattern of a NSCO thin film grown onto LAO (001). In this case, as stated above, the lattice parameters of the substrate and those of the cobaltite are closer with respect to the STO. Pattern in the main Figure shows the presence of only two peaks exactly located at the position of the (001) and (002) reflections of the LAO substrate. In the inset of the Figure it is presented the comparison of the diffraction pattern of the NSCO film on LAO (dashed line) and of the substrate (solid line). As can be appreciated, the only difference between the two patterns is the lack, for the NSCO film, of the clear $K\alpha_1/K\alpha_2$ separation visible in the LAO pattern. This means that the film has grown with lattice constants so close to those of the LAO substrates that is not possible to discriminate between them. This is also due to the FWHM of the films being of the order of 0.3° with respect to the 0.05° of the single crystal peaks.

## Conclusion

In this paper we reported, to the best of our knowledge, the first deposition of highly oriented thin films (with thickness of about 90 nm) of $NdCoO_3$ and $Nd_{0.8}Sr_{0.2}CoO_3$ cobaltites on single-crystalline STO and LAO substrates.

Our investigation has shown that highly oriented single phase thin films of NCO and NSCO can be successfully deposited by means of rf-sputtering if the substrates is heated at high temperatures (700°C); lower substrate temperature has shown to lead to multi-phase materials with a low crystallinity degree .

Post-deposition annealing treatments in oxygen are efficient in increasing the oxygen content of the samples, as witnessed by the lattice constant reduction, but do not significantly enhance the film crystallinity. LAO substrate showed to give origin to a prefect match of the out-of-plane lattice constant of the NSCO target material.

## Acknowledgement

Financial support from the Italian Ministry of Scientific Research (MIUR) by PRIN Projects (2004) is gratefully acknowledged. One of us (L.M.) gratefully acknowledges the financial support of the "Accademia Nazionale dei Lincei".

# Figures Captions

**Figure 1** – Rietveld refined pattern of NCO. Grey squares represent the experimental pattern, black line the calculated one while vertical bars at the bottom of the pattern are the Bragg peaks position. Horizontal black line shows the difference between the calculated and experimental patterns.

**Figure 2** – X-ray reflectivity spectrum of NCO deposited for 30' on STO (001). The inset highlights the region at low angle.

**Figure 3** – XRD pattern for as-deposited (solid line) and annealed (short-dashed line) NSCO on STO. Vertical black lines indicate the position of orthorhombic NSCO peaks, while the grey ones refer to the STO. Inset: enlargement of a small region of the pattern around the main peak.

**Figure 4** –XRD pattern of NSCO on STO with Miller indexes for the two possible epitaxial orientations.

**Figure 5** – XRD pattern for NCO on STO (001). Inset: comparison between NCO (short-dashed line) and NSCO (solid line) XRD around the main peak.

**Figure 6** – XRD pattern for NCO deposited at 400°C. Asterisks mark the extra-peaks not directly related to the cobaltite.

**Figure 7** – XRD pattern for NSCO on LAO (001). Inset: comparison between the XRD pattern for the LAO substrate (solid line) and NSCO (short-dashed line).

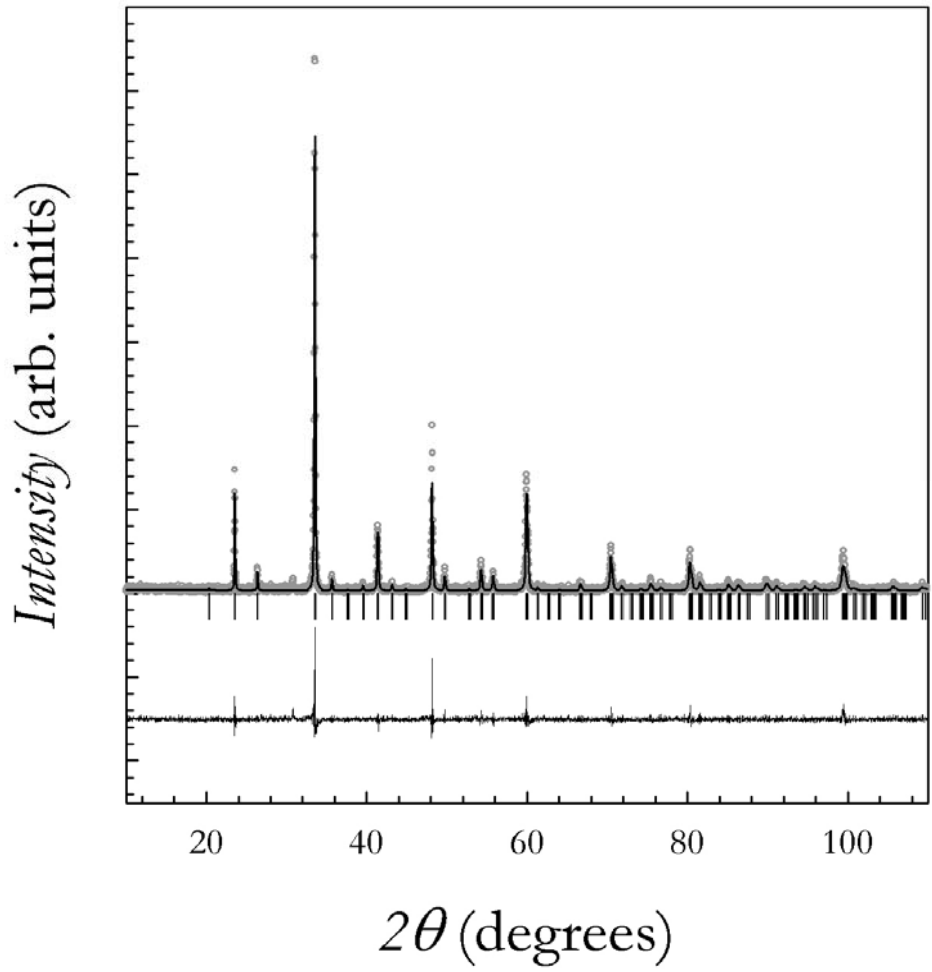

**Figure 1**

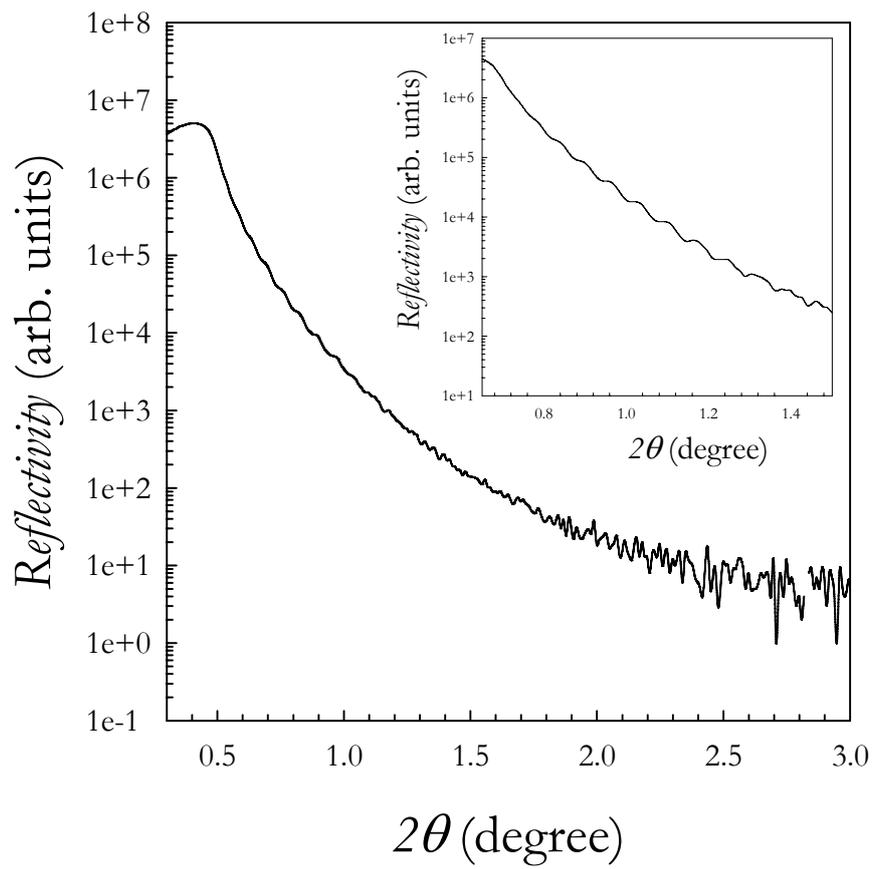

**Figure 2**

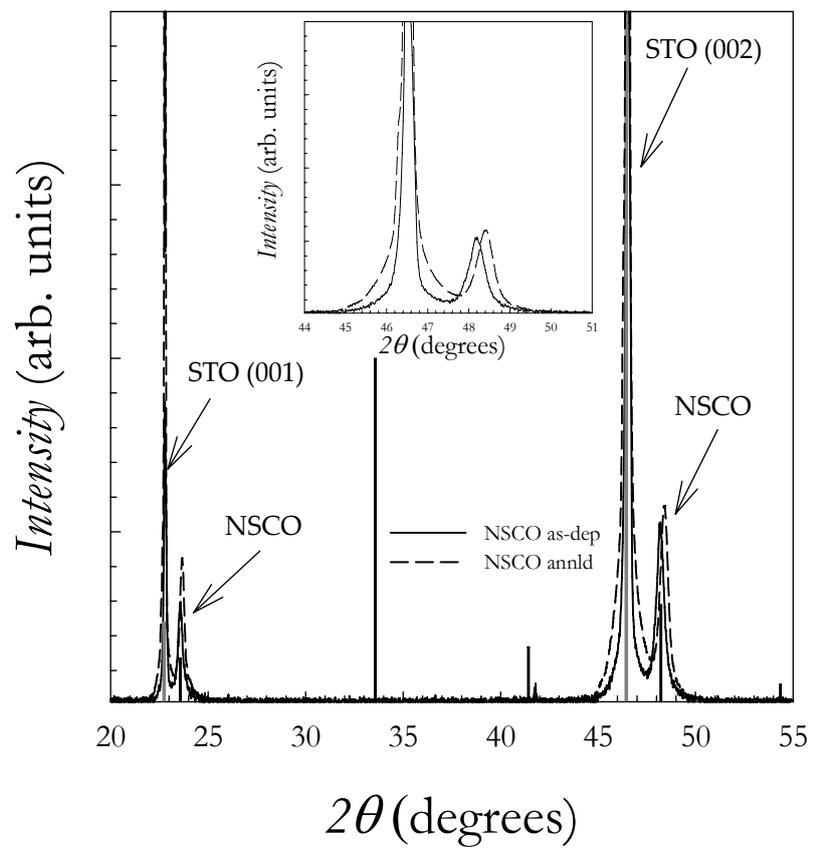

**Figure 3**

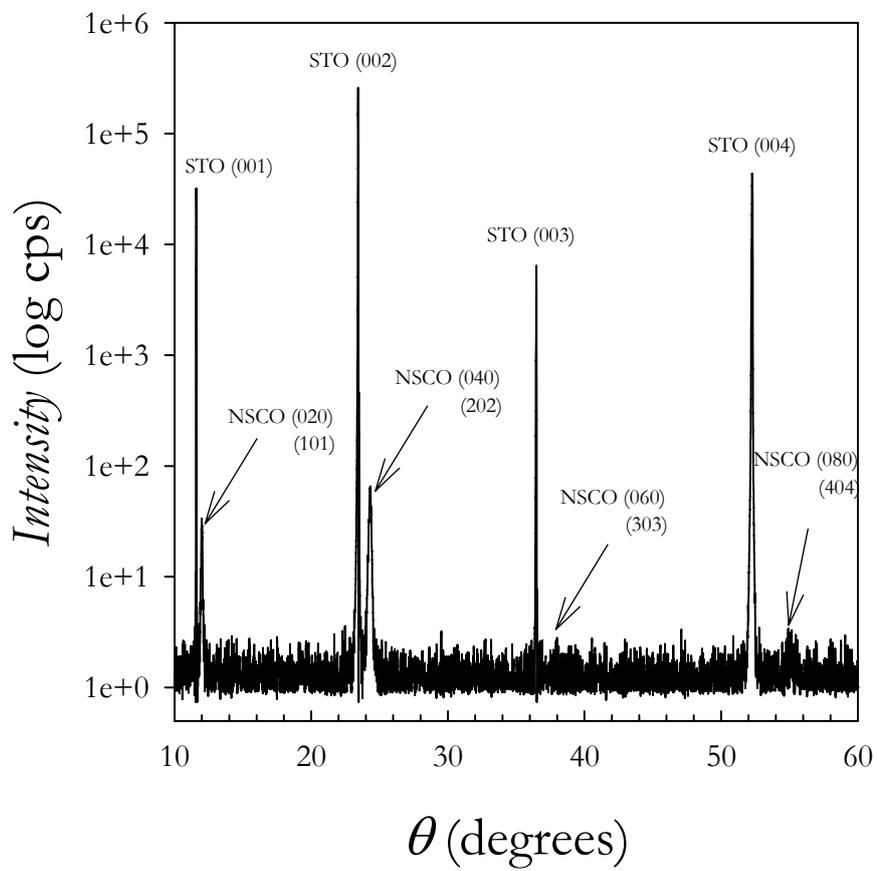

**Figure 4**

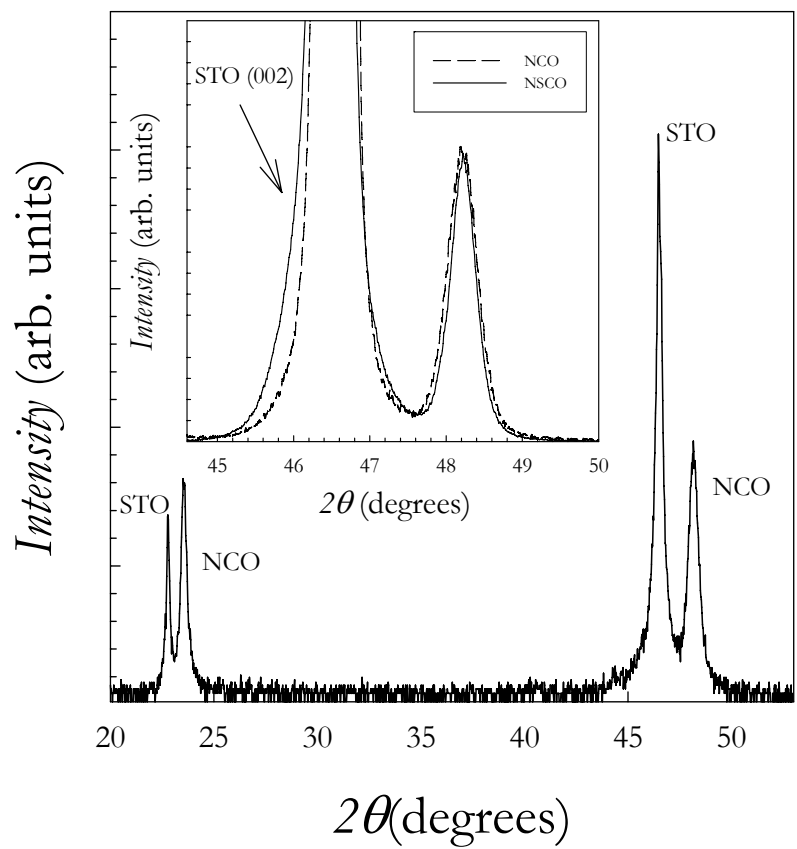

**Figure 5**

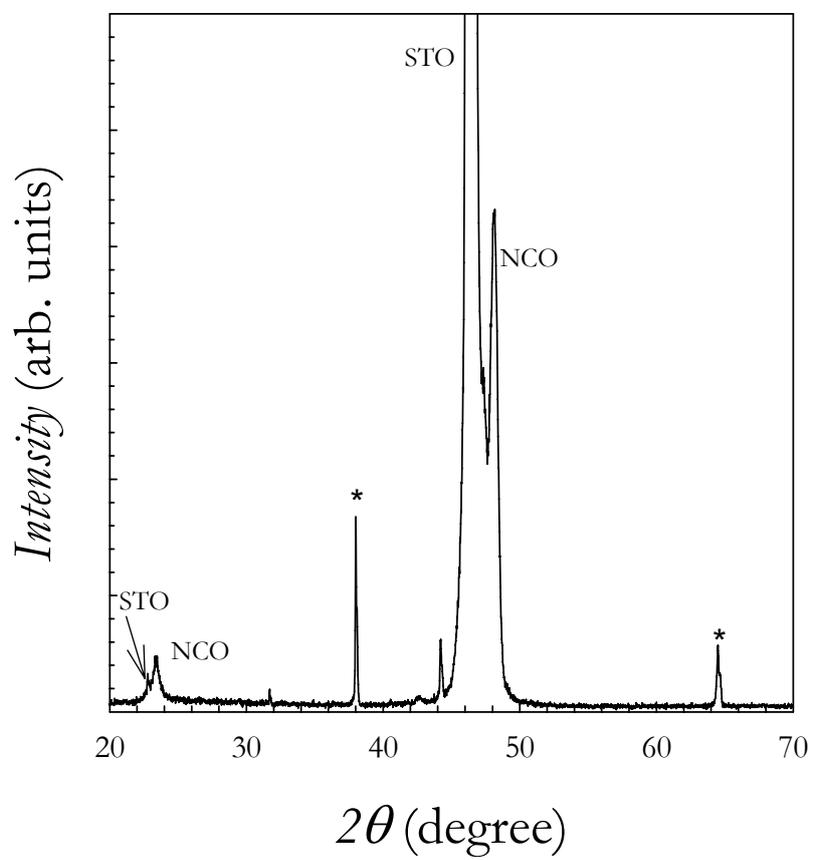

**Figure 6**

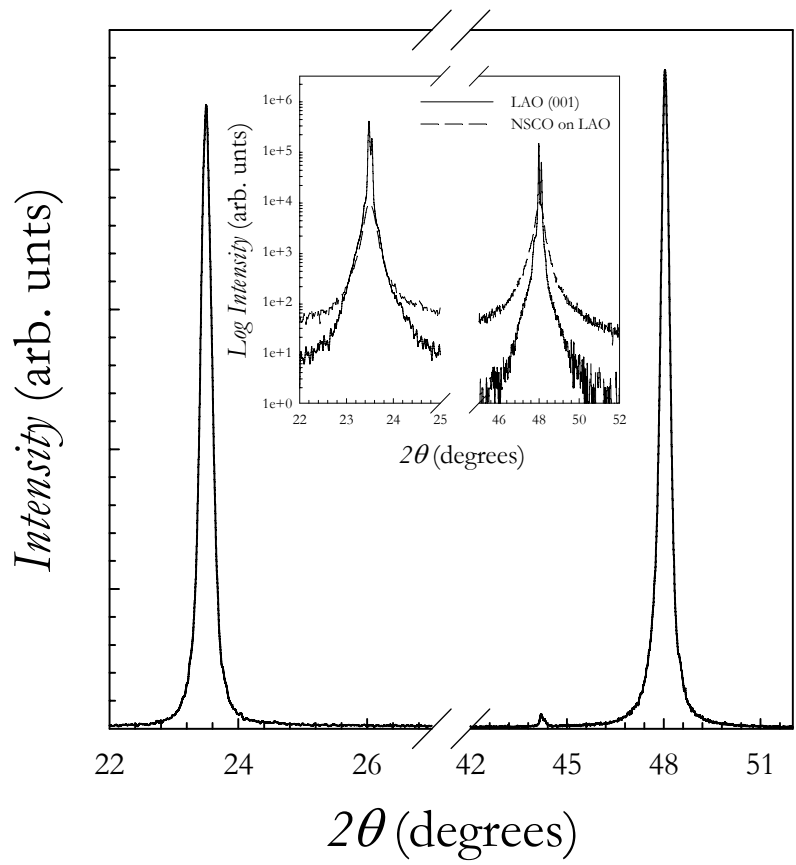

**Figure 7**